\documentclass[12pt,english,aps,prl,twocolumn,showpacs,superscriptaddress,
footinbib,reprint,noshowpacs]{revtex4-1}
\usepackage{graphicx}
\def\arXiv#1{\href{http://arxiv.org/abs/#1}{arXiv:#1}}
\usepackage{amsmath}
\usepackage{comment}
\usepackage{amssymb}

\RequirePackage{amsmath,amssymb,amsthm,graphicx,mathrsfs,url}
\RequirePackage[usenames,dvipsnames]{color}
\RequirePackage[colorlinks=true,linkcolor=Red,citecolor=Green]{hyperref}
\RequirePackage{amsxtra}

\def\indic{\operatorname{1\hskip-2.75pt\relax l}}
\usepackage{wasysym}

\usepackage{subcaption} 
\def\smallsection#1{\smallskip\noindent\textbf{#1}.}
\usepackage{float}
\begin{document}

\title{Honeycomb structures in magnetic fields}

\author{Simon Becker}
\email{simon.becker@damtp.cam.ac.uk}
\affiliation{Department of Applied Mathematics and Theoretical Physics, University of Cambridge, Wilberforce Road, Cambridge, CB3 0WA, United Kingdom.}

\author{Rui Han}
\email{rui.han@math.gatech.edu}
\affiliation{School of Mathematics, Georgia Institute of Technology, Atlanta, Georgia  30318, USA.}

\author{Svetlana Jitomirskaya}
\email{szhitomi@math.uci.edu}
\address{Department of Mathematics, University of California, Irvine, CA 92697, USA.}

\author{Maciej Zworski}
\email{zworski@math.berkeley.edu}
\affiliation{Department of Mathematics, University of California,
Berkeley, CA 94720, USA.}

\begin{abstract}
We consider
% tight-binding 
reduced-dimensionality models of honeycomb lattices in magnetic
fields and report results about the spectrum, the density of states,
self-similarity, and metal/insulator transitions under disorder. We perform a spectral analysis by which we
discover a fractal Cantor spectrum for irrational magnetic flux
through a honeycomb, prove the existence of zero energy
Dirac cones for each rational flux, obtain an explicit expansion of the density of states near the conical points, and show the existence of mobility edges under Anderson-type disorder. Our results give a precise description of
de Haas-van Alphen and Quantum Hall effects, and provide
quantitative estimates on transport properties. In particular, our
findings explain experimentally observed asymmetry phenomena by going beyond the perfect cone approximation.
\end{abstract} 

\maketitle
\makeatother

% \smallsection{Introduction of the model and main results} 
Reduced-dimensionality models are of central importance in condensed matter physics as they are often analytically solvable and allow for a qualitative description of material properties.  

A phenomenologically rich class is composed of tight-binding and
infinite-contrast models on periodic lattices with constant magnetic
fields. Following a thorough study of these models over the past
fourty years, rigorous results on the fractal spectrum on the $\mathbb
Z^2$-lattice \emph{(Harper's model)}
\cite{H1}-\cite{H13}%H2,H3,H4,H5,H6,H7,H8,H9,H10,}
, the location of the low-lying spectrum \cite{HS0,HS1,HS20,HS2}, and the disordered model \emph{(Anderson model)} \cite{GKS,FS83,Anders} have been obtained.
 
\begin{figure}%[t!]
% \centering
\captionsetup{justification=raggedright,font=small,
singlelinecheck=false
}
%  \begin{subfigure}{0.18\textwidth}
   \includegraphics[width=5.5cm]{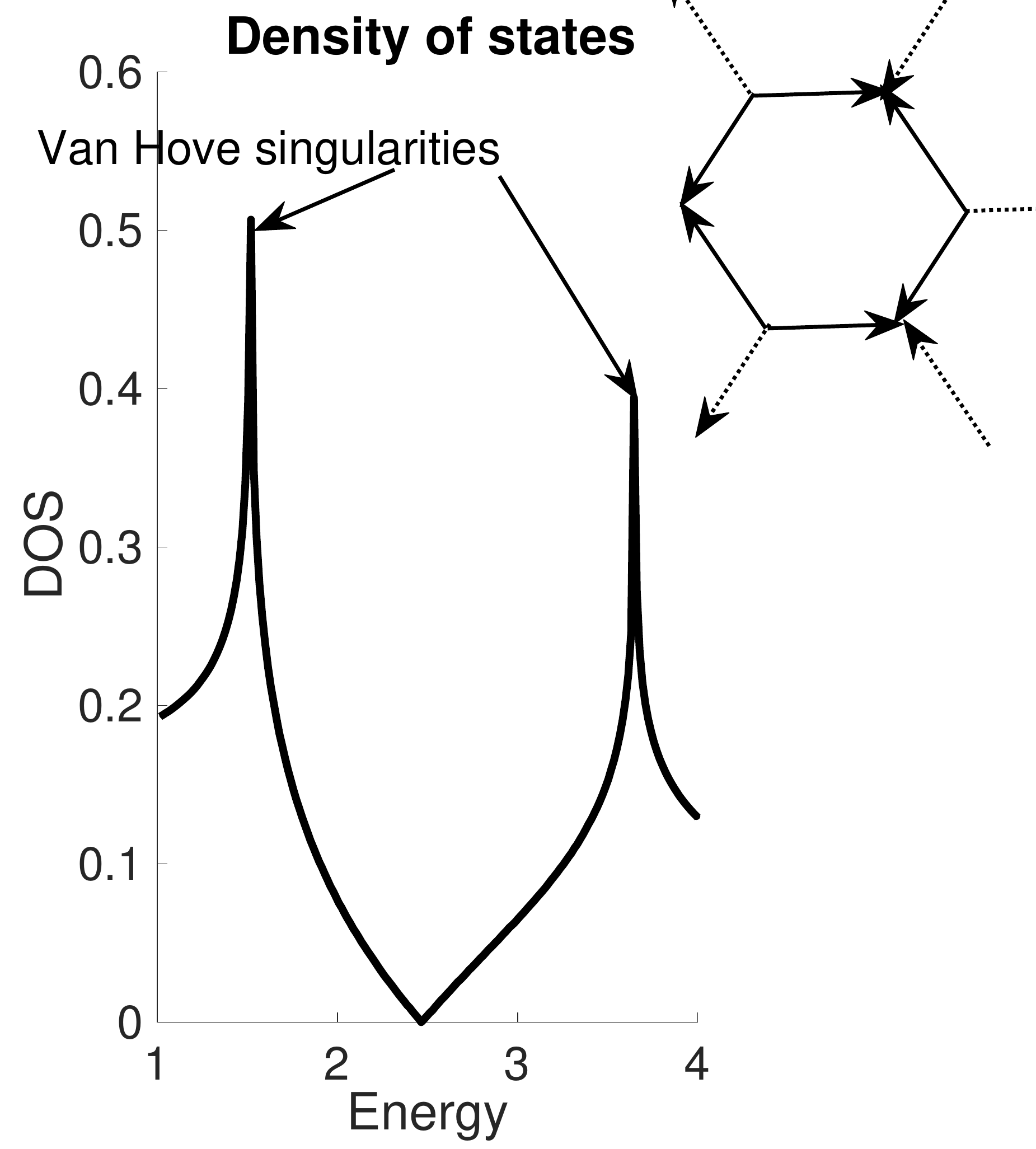}
    %\caption{The DOS of $H^{B=0}$ \eqref{eq:HB} and $V=0$ on the first band $[0,\pi^2]$.}
%  \end{subfigure}
%\qquad \
%  \begin{subfigure}{0.18\textwidth}
%    \includegraphics[width=3.9cm]{hari.pdf}
%    \caption{Experimental plot of the DOS for molecular graphene  \cite{hari}.}
%  \end{subfigure}
  \caption{\label{Fig:DOS} 
  The DOS of the quantum graph model without magnetic of field
  obtained using  $H^{B=0}$ \eqref{eq:HB} and $V=0$ on the first band $[0,\pi^2]$. The comparison with \cite{hari} Fig.1 shows a good agreement of the model with molecular graphene experiments.}
\end{figure}

\begin{figure}[t!]
  %\centering
   \includegraphics[height=7cm,width=7.6cm]{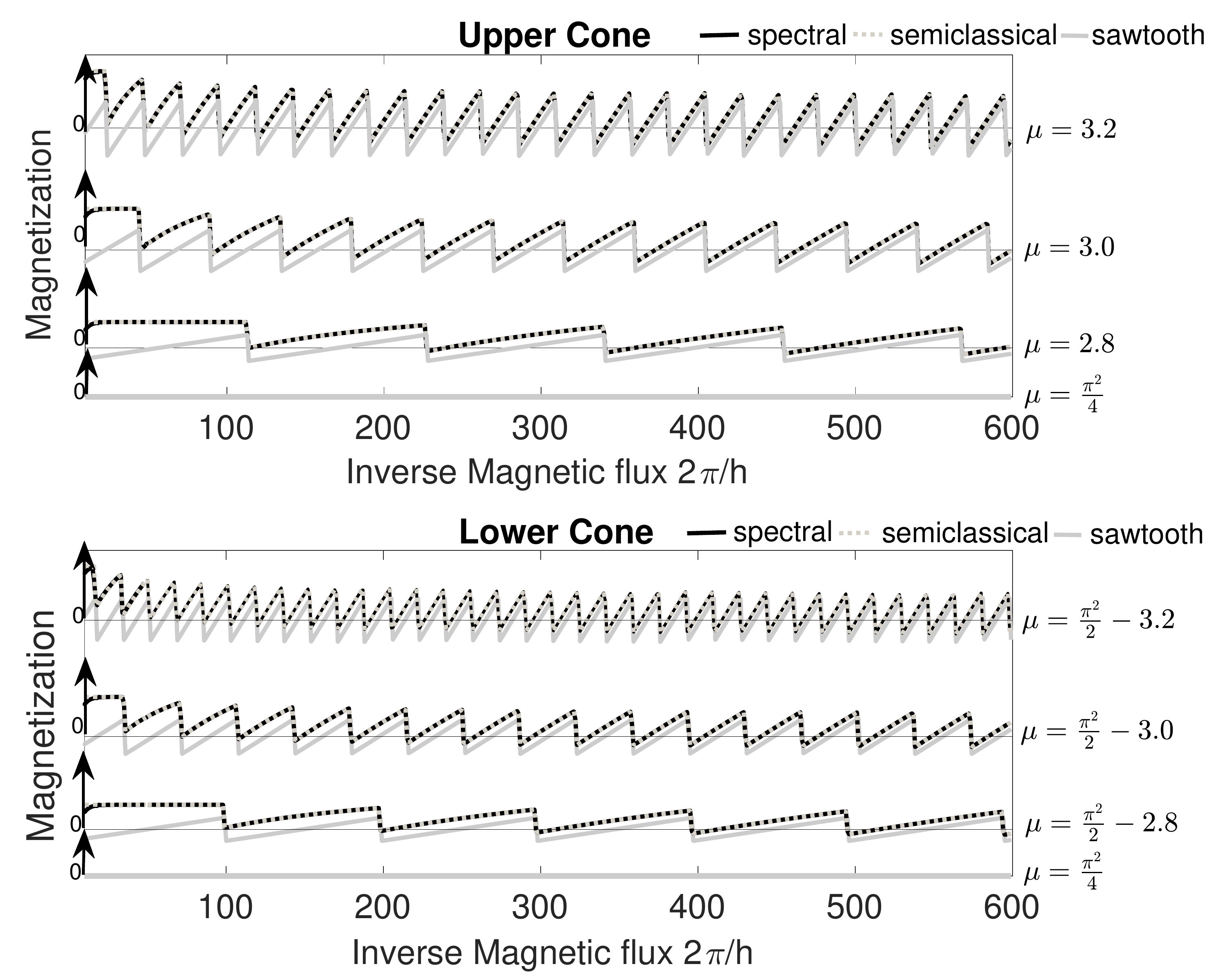} 
   \captionsetup{justification=raggedright,font=small,
singlelinecheck=false
}
\caption{\label{Fig:cut-off} Magnetization  
for different chemical potentials {\em above/below} the Dirac energy $\mu=\tfrac{\pi^2}{4}$, with $V \equiv 0.$ } 
\end{figure}

\begin{figure}[t!]
 \includegraphics[ trim={3cm 0 0 0},height=5.5cm, width=10.0cm]{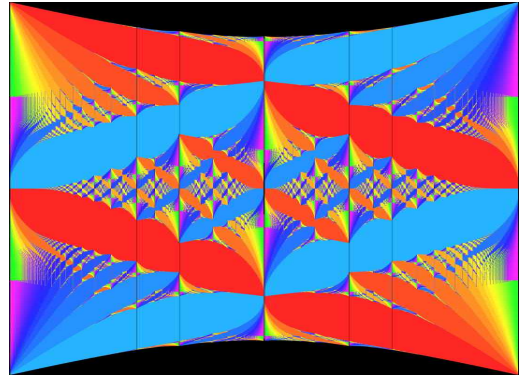}
 \captionsetup{justification=raggedright,
singlelinecheck=false,font=small
}
 \caption{Hofstadter butterfly on honeycomb lattice. Different colours correspond to different Hall conductivities.\label{fig:Hofstadter}}
\end{figure}

The purpose of this letter is to report rigorous results on tight-binding and infinite-contrast models for honeycomb structures in constant magnetic fields and emphasize new theoretical approaches.
We consider the Hamiltonian, $H^B$, with constant magnetic field $B$,  defined on edges $e$ of a honeycomb graph $\Lambda$ by 
\begin{equation}
\label{eq:HB}
H_e^B:=(-i \partial_x-A_e )^2 + V_e, \ \  e \simeq (0,1).
\end{equation}
We assume Kirchhoff boundary conditions \cite{KP} at the vertices and
that  $V_e$ is symmetric with respect to the centre of the edge
$e$. Such Hamiltonians are called \emph{quantum graphs} or \emph{wave
  guides} \cite{KS,KP} and %generalize 
represent the 
%tight-binding 
infinite-contrast limits \cite{HKL} of continuous Schr\"odinger operators on $\mathbb R^2$ with honeycomb lattice potential \cite{FW,FLW,D,D2}. 
Apart from the interest in such models as %tight-binding
 limits (see \cite{BGP, AJ} for related work on Harper's model), quantum graphs are natural models for \emph{molecular graphene} \cite{hari,PoMa} and \emph{wave guides} \cite{KS2}, see Fig.\ref{Fig:DOS}. In this letter, we explain three key physical phenomena:
 \begin{itemize}
 \item \emph{Spectral theoretic}: We provide a  full  spectral analysis. In particular, we show that for irrational magnetic flux through a single honeycomb, the spectrum is a Cantor set of %Lebesgue 
 measure zero and Hausdorff dimension at most $1/2$ \cite{BHJ17,JK2}. 
 \item \emph{Semiclassical}: We derive a semiclassical expansion for the density of states (DOS) supported on geometric Landau levels and relate it to the Shubnikov-de Haas, de Haas-van
Alphen, and Quantum Hall effects \cite{BZ,B}.  We note a remarkable
agreement of the asymptotic result with exact spectral calculations,
see Fig.\ref{Fig:cut-off}. This analysis holds near each
conical singularity, which we show to exist for each rational flux,
thus providing 
%and provide 
a foundation for self-similarity appearing in Fig.\ref{fig:Hofstadter}.
 \item \emph{Dynamical}: For Anderson-type potentials in weak magnetic fields under weak disorder, we identify insulating regions away from the Landau levels in which Anderson localization occurs, and regions of metallic transport close to the Landau levels \cite{B}. 
 \end{itemize}
For vector potentials $A(x)=A_1(x)dx_1+A_2(x)dx_2 $ the scalar potential $A_e$ in \eqref{eq:HB} along edges
$e$ is $A_{e}(x) := A(x)\left(e_1 \partial_1 + e_2 \partial_2 \right)$ \cite{KS}.
%The integrated vector potentials are defined as $\beta_{e}:=\int_{e} A_{e}(x) dx.$
%where $\kappa_{e}$ is the canonical chart of the edge $e$, see (\ref{chart}).
The magnetic flux $h:= \int_{\varhexagon} dA$ through each honeycomb $\varhexagon$ of the lattice 
is taken to be constant.
The Hamiltonian $H^B$ on the graph can be identified with the standard tight-binding operator $t^h$ on the honeycomb lattice:
After a simple change of geometry, the tight-binding operator $t^h$ is
\begin{equation}
\label{eq:Q}
t^h= \tfrac{1}{3} \left(\begin{matrix} 0 && 1+\tau^0+\tau^1 \\ \left(1 +\tau^0+\tau^1 \right)^* && 0 \end{matrix} \right) \text{ on } \mathbb{Z}^2
\end{equation}
with the magnetic translations
$\tau^0(r)(\gamma):= r(\gamma_1-1,\gamma_2), \tau^1(r)(\gamma):= e^{i h \gamma_1}r(\gamma_1,\gamma_2-1)$ for $ \gamma \in \mathbb{Z}^2,$ and $r \in \ell^2(\mathbb{Z}^2;\mathbb{C}).$ 
Solving $-y_{\lambda}''(x)+V_ey_{\lambda}(x)=\lambda y_{\lambda}(x)$,  $y_{\lambda}(0)=1,y_{\lambda}'(0)=0,$ we put $\Delta(\lambda):=y_{\lambda}(1)$. Then 
 $\lambda \in %\mathbb R$ is in the spectrum $
 \operatorname{Spec}(H^B) \setminus \operatorname{Spec}(H^D) $ 
 ($H^D$ is the operator \eqref{eq:HB} on a single edge with Dirichlet boundary conditions) if and only if $\Delta(\lambda) \in \operatorname{Spec}(t^h)$ -- see \cite{BGP,BHJ17,P,P13,P14}. Since $\left\lVert t^h \right\rVert <1$ \cite{BGP,BHJ17} for non-trivial magnetic flux $h \notin 2\pi \mathbb Z$ the spectrum of $H^B$ decomposes into the disjoint union of continuous spectrum $\Delta^{-1}(\operatorname{Spec}(t^h))$ and infinitely degenerate eigenvalues $\lambda \in \operatorname{Spec}(H^D) $, see \cite{BHJ17}.

\begin{figure}
{\includegraphics[width=7.5cm]{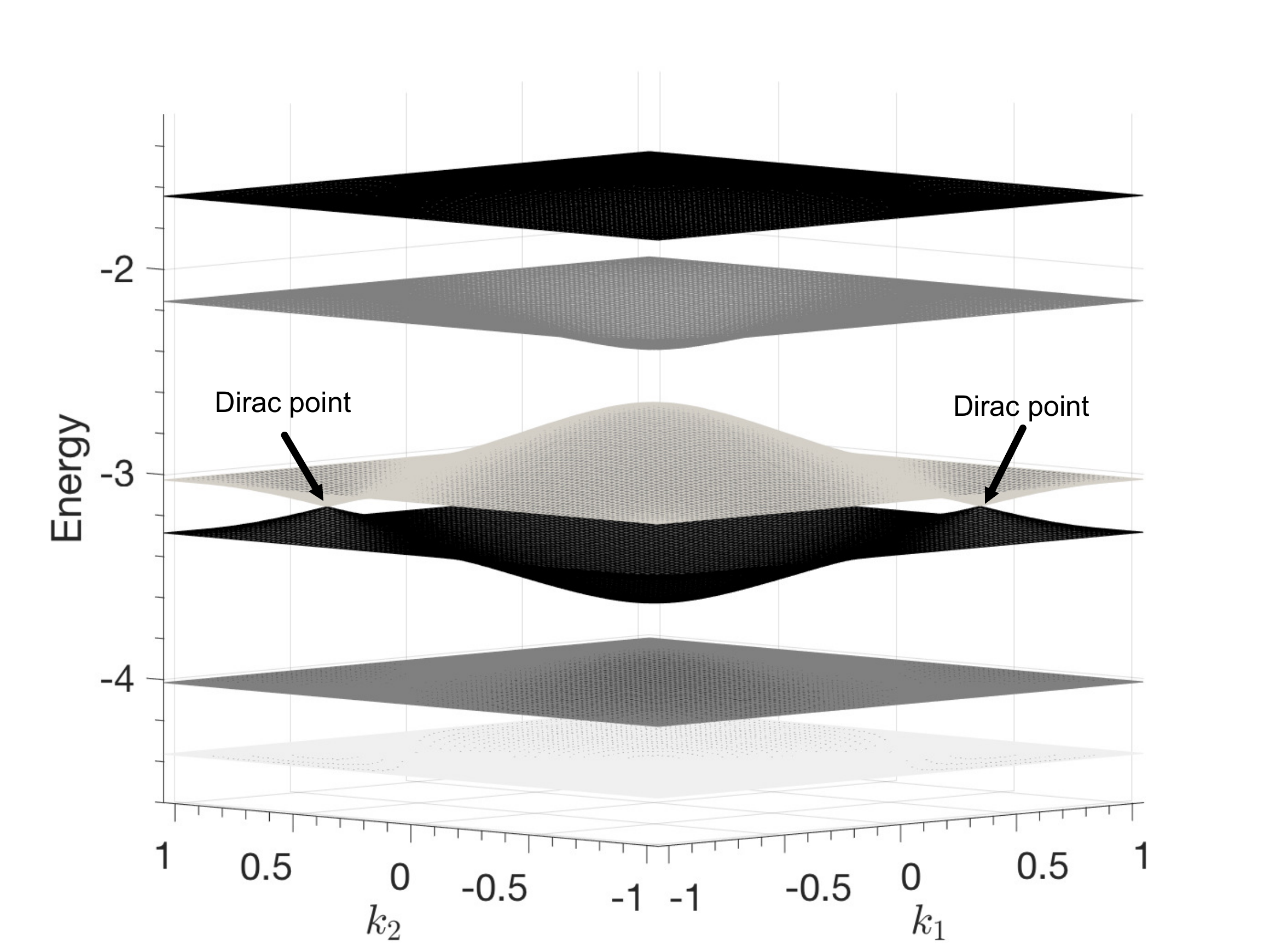}} 
\captionsetup{justification=raggedright,
singlelinecheck=false,font=small
}
\caption{We show that Dirac points persist under rational flux $\frac{h}{2\pi} \in \mathbb Q$. Here, $\frac{h}{2\pi}=\frac{1}{3}$ with Mathieu potential $V(x)=20\cos(2\pi x)$.\label{Fig:notouchingbands} }
\end{figure}

\begin{figure}[t!]
\captionsetup{justification=raggedright,font=small,
singlelinecheck=false
}
\center\includegraphics[width=7.5cm]{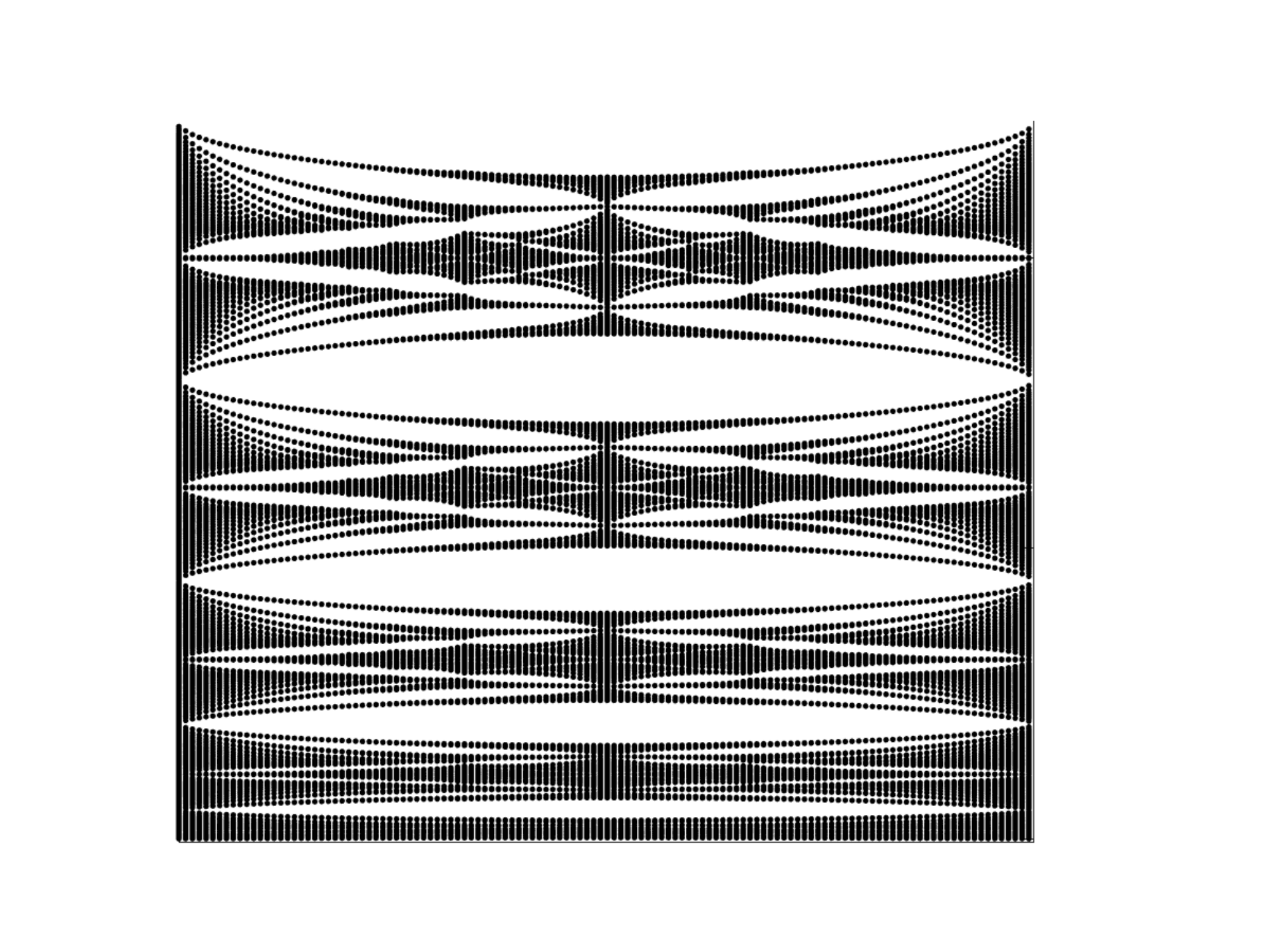} 
\caption{\label{Fig:hof} Hofstadter butterfly on honeycomb lattice. The spectrum of $H^B$ is plotted as a function of the magnetic flux $h \in [0,2\pi]$. }
\end{figure}

\smallsection{Cantor spectrum:} The fractal structure of magnetic electron spectra was
first predicted by Azbel \cite{Az} and then numerically confirmed by
Hofstadter \cite{Ho} for Harper's model, see Fig.\ref{Fig:hof}. Verifying this experimentally is difficult
as the smallness of the cell requires extraordinarily strong magnetic
fields to obtain observable magnetic flux. Only recently, self-similar
structures in the electron spectrum of graphene have been observed
\cite{Ch,C14,Ga,Gor}. Earlier experiments involved modeling of 
periodic structures by microwaves \cite{kuhl}. Here, we first assume that the normalized magnetic flux $\frac{h}{2\pi}=\tfrac{p}{q}$ is rational, as then the Floquet-Bloch theory implies that the spectrum of \eqref{eq:Q} has band structure, see Fig.\ref{Fig:notouchingbands}. We can then express the spectrum of \eqref{eq:Q} using a 1D-Jacobi operator with quasi-momentum $k \in \mathbb T_1^*=[0,2\pi]$
\begin{equation}
\begin{split}
\label{defH}
(Ju)_m&=\left(1+e^{ i \left(k+m h \right)} \right) u_{m+1} +2 \cos \left(k+m h\right)u_m\\
&\quad +\left(1+e^{ i \left(k+(m-1) h \right)} \right)u_{m-1}
\end{split}
\end{equation}
and from the study of such singular Jacobi operators \cite[Lemma $4.3$]{BHJ17} we estimate the Lebesgue measure % $\vert \bullet \vert$
\begin{equation}
\label{eq:measure}
\vert \operatorname{Spec}(t^h) \vert \lesssim q^{-1/2}.
\end{equation}
The spectrum of \eqref{eq:Q} is continuous (in Hausdorff distance $d_H$) with respect to the magnetic flux \cite[Lemma $6.2$]{BHJ17} 
\begin{equation}
\label{eq:cont}
d_{H}\left(\operatorname{Spec}\left(t^h\right),\operatorname{Spec}\left(t^{h'}\right)\right) \lesssim \left\vert h-h'\right\vert^{1/4}.
\end{equation}
However, the spectral nature for irrational fluxes changes dramatically, see \cite[Thm. $3$]{BHJ17}. If $\tfrac{h}{2\pi}$ is irrational, the spectrum of \eqref{eq:Q}, and the continuous spectrum of \eqref{eq:HB}, is a fully disconnected and nowhere dense set without isolated points of measure zero with Hausdorff dimension at most $\frac12$ \cite[Thm.$1.5$]{JK2}.  
For irrational fluxes $\tfrac{h}{2\pi}$ with unbounded continued
fraction expansion, the Lebesgue measure of the spectrum vanishes by
combining estimate \eqref{eq:measure} and the continuity estimate
\eqref{eq:cont}. Since the spectrum is always closed and, as can be
shown, it has no isolated points, this implies Cantor-type
spectrum. Using Kotani's theory, the Cantor structure of the spectrum can also be shown to hold for all irrational fluxes $\tfrac{h}{2\pi}.$ The bound on the Hausdorff dimensions follows from an almost Lipschitz continuity estimate on the spectrum of singular quasiperiodic Jacobi operators obtained in \cite{JK2}.

%If $h/(2\pi)$ is irrational, the spectrum of \eqref{eq:Q}, and the continuous spectrum of \eqref{eq:HB}, is a fully disconnected and nowhere dense set without isolated points of measure zero with Hausdorff dimension at most $1/2$ for generic $h$. \newline 
%For irrational fluxes $h/(2\pi)$ with unbounded continued fraction expansion, the Lebesgue measure of the spectrum vanishes by combining estimate \eqref{eq:measure} and the continuity estimate \eqref{eq:cont}. Since the spectrum is always closed and it can be argued that $\eqref{eq:Q}$ has no point spectrum, and thus no isolated points, this implies Cantor-type spectrum. Using Kontani theory, the Cantor spectrum can then be established for all irrational fluxes $h/(2\pi).$ The bound on the Hausdorff dimensions follows by approximating the spectrum of \eqref{eq:Q} for irrational fluxes $h/(2\pi)$ by coverings of the spectrum with rational flux, similar to \cite{Last}, and using the continuity estimate to control the error. 

\begin{figure}[t!]
\captionsetup{justification=raggedright,font=small,
singlelinecheck=false
}
\includegraphics[width=8.6cm]{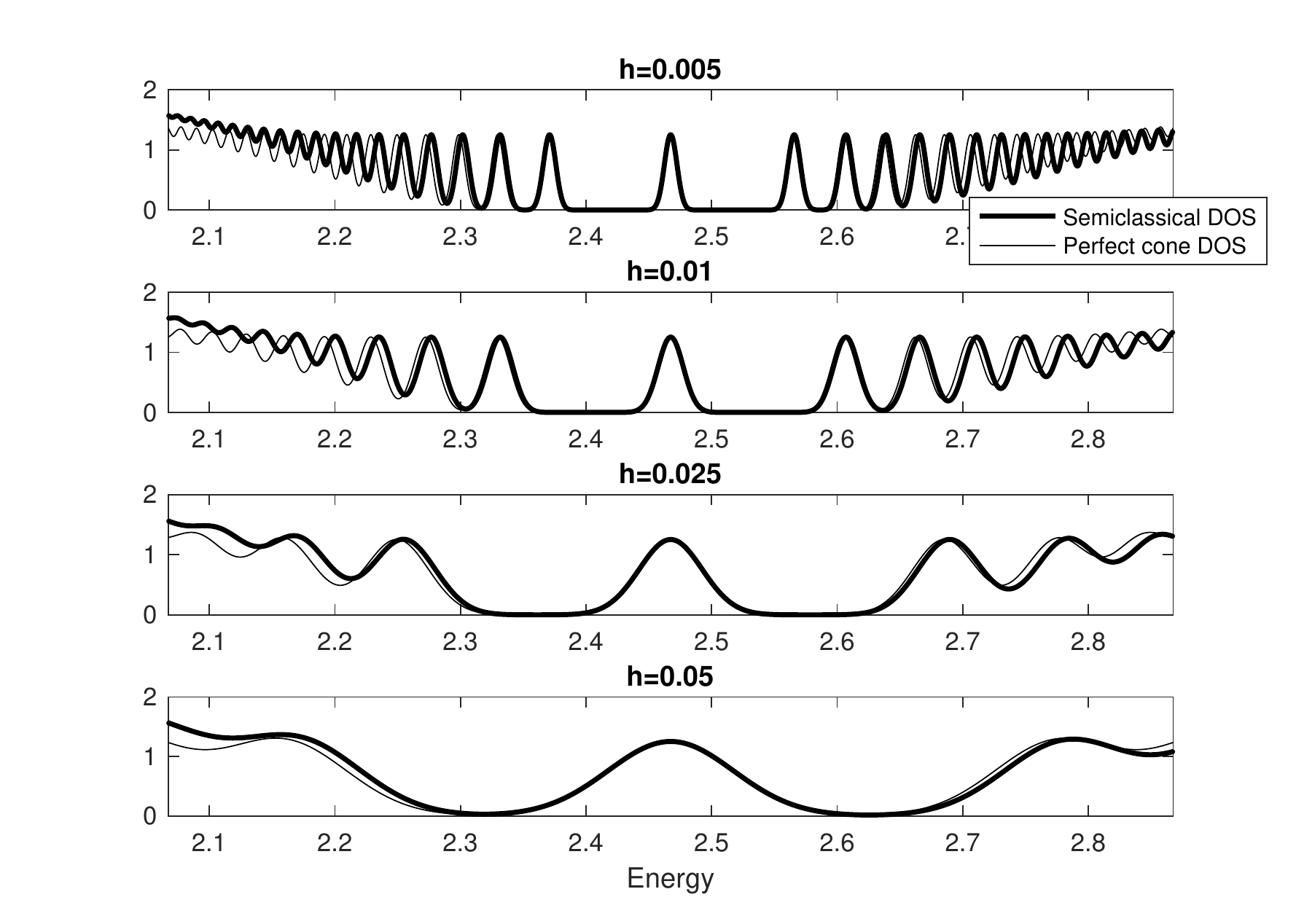} 
\caption{\label{fig:DOS4} \emph{Shubnikov-de Haas oscillations} of $ \mu \mapsto \rho_{H^B} ( \exp ( ( \bullet - \mu)^2 /2 \sigma^2 ) / \sqrt{ 2 \pi} \sigma ) $ for different values of $ h$. We note the asymmetry when compared to the 
DOS assuming perfect cones.}
\end{figure}

\smallsection{Semiclassical analysis of the DOS:} The density of states is a \emph{generalized function} $\rho_{H^B}$ defined in terms of the regularized trace 
\[\widetilde{\operatorname{tr}}(f(H^B)) = \lim_{r \rightarrow \infty} \frac{ \operatorname{tr}\indic_{B_r(0)} f(H^B) }{\vert B_r(0) \vert}=\int_{\mathbb R} f(x) \rho_{H^B}(x) \ dx\] 
where $ B_r(0)$ is the % volume of the
 ball of radius $r$, see Figs. \@\ref{Fig:DOS} and
 \ref{fig:DOS4}. 
By spectral equivalence of \eqref{eq:HB} and \eqref{eq:Q}, 
for energies close to the Dirac point energy it suffices to analyze the DOS of $t^h$.  The magnetic translations in \eqref{eq:Q} satisfy the Weyl commutation relations $\tau^1 \tau^0 = e^{i h} \tau^1 \tau^2$ and
the same commutation relation is obtained for 
$D_x := -i \frac{\partial}{\partial x}$ by 
$ e^{ ih D_x } e^{ix} = e^{ i h} e^{ix} e^{ i h D_x }  $
%$ e^{ih D_x} e^{ix} = e^{i h} e^{ix} e^{i h D_x}$ 
where $e^{i h D_x} = \operatorname{Op}^{\text{w}}_{h}(e^{i\xi})$ is the Weyl quantization of the symbol $e^{i \xi}$ \cite{Zw}.
This different representation reduces the analysis of the DOS of \eqref{eq:Q} to the study of the DOS of the operator
\begin{equation}
\label{eq:Diffop}
 \tfrac{1}{3} \left(\begin{matrix} 0 && 1+e^{ix}+\operatorname{Op}^{\text{w}}_{h}(e^{i\xi}) \\ 1+e^{-ix}+\operatorname{Op}^{\text{w}}_{h}(e^{-i\xi}) && 0 \end{matrix} \right).
\end{equation} 
Through a symplectic change of variables, $ y = a ( x + \xi ) $, $ \eta = b \left( \xi - x \pm \tfrac{4 \pi } 3 \right)$,  
($a = \pm 2^{-\frac12} 3^{-\frac14} $,  $ b = \pm 2^{-\frac12} 3^{\frac14} 
$)
one finds that at the Dirac points we have
\begin{equation}
\label{eq:norfo}
\begin{split}  1 + e^{ i x } + e^{ i \xi } & = 
%\frac{ \sqrt 3 }{2} a \eta - i \frac12 b
%y + \mathcal O ( y^2 + \eta^2 ) = 
c ( \eta \mp i y )  + \mathcal O ( y^2 + \eta^2 ) , \\
1 + e^{ -i x } + e^{ - i \xi } & = c ( \eta \pm i y )  + \mathcal O ( y^2 + \eta^2 ), 
\end{split}
\end{equation}
$ c = 3^{\frac14} 2^{-\frac12} $. 
Classical-quantum correspondence implies that by the symplectic change of variables (\emph{classical}), operator \eqref{eq:Diffop} is (micro)-locally equivalent (\emph{quantum}) to the operator $ \tfrac{c}{3} \left(\begin{matrix} 0 && a_{\pm}^* \\ a_{\pm} && 0 \end{matrix} \right)$ quantized in new variables $a_{\pm}:=y \pm i hD_{y}.$
The spectrum of this operator can be explicitly expressed through the quantum harmonic oscillator. 
By making these steps precise and taking higher order contributions of
the geometry in \eqref{eq:norfo} into account, it is possible to show
the %following 
semiclassical Bohr-Sommerfeld description of the DOS with precise error control \cite[Thm. $1$]{BZ}:
If $I\subset \Delta^{-1}  (-\delta,\delta)$, with $\delta>0$ small, then
\begin{equation}
\begin{split}
\label{eq:Landau}\begin{gathered}
\widetilde{ \operatorname{tr}} f (  H^B) = \tfrac{2h}{3 \sqrt{3}\pi}  \sum_{ n \in \mathbb Z^2 } 
f ( z_{n} ( h  ) )+ \mathcal O (\vert  f \vert_{ C^\alpha } h^\infty) \  \\ 
\Delta(z_{n} ( h ) )= \kappa ( n h , h ),  
\ \ \alpha > 0 , 
 \end{gathered}
\end{split}
\end{equation}
$\kappa ( n h , h )$ are the solutions to the Bohr-Sommerfeld condition $F ( \kappa ( \zeta, h )^2 , h  ) = |\zeta| + \mathcal O 
(h ^\infty)$ with the expansion
\begin{equation}
\label{eq:g2F2}  
\begin{gathered} 
F ( s , h ) \sim  F_0 ( s ) + \sum_{j=2}^{\infty} h^j F_j ( s ), \ F_0 ( s  ) =  \tfrac1{4 \pi}{\int_{ \gamma_s } \xi dx}  , \\
\gamma_s = \left\{ ( x, \xi) \in \mathbb{T}^2_*: \tfrac{| 1 + e^{ix } + e^{i\xi} |^2}{9} = s \right\}, \ \ F_j ( 0 ) = 0. 
\end{gathered} 
\end{equation}
In particular, we show that $F_1(s)=0$ for all such $2 \times 2$ operators with only off-diagonal contributions.
Writing  $ g ( x ) = F_0 ( \Delta ( x ) ^2 )$, we obtain a leading order approximation of Landau levels
\begin{equation}
\label{eq:appz}
z_{ \pm |n|}^{(1)} ( h ) = g_{\pm}^{-1} ( |n| h ) , \ \  z_0^{(1)} ( h ) = 0.
\end{equation}
In \eqref{eq:Landau}, $ \vert f \vert_{ C^\alpha } := 
\sup_{x} |f ( x) | + \sup_{ x \neq y } \tfrac{|f ( x ) - f ( y ) |}{|x-y|^{\alpha } } $: it is essential to allow non-smooth test functions $ f $ in view of applications to magnetic oscillations. (See for instance
\cite{SU} for a physics perspective on semiclassical approximation in this setting.) 

Fig.\ref{fig:DOS4} shows that Landau levels, and thus the DOS are non-symmetric with respect to the Dirac point energy.  We compare this with the symmetric leading order \emph{(perfect cone)} approximation of phase space area $g_c(x)= {(x-z_D)^2}/{v_F^2}$ with Fermi velocity $v_F = 3^{-3/4} \Delta'(z_D)^{-1}$ and Dirac point energy $z_D \in \Delta^{-1}(0).$
\eqref{eq:Landau} also explains \emph{de-Haas van Alphen oscillations}. To formulate it we introduce the \emph{grand-canonical potential} with inverse temperature $\beta$ and chemical potential $\mu$:
\begin{equation}
\label{eq:Omy}
\begin{gathered}
\Omega_\beta ( \mu , h ) := \rho_B ( \eta ( \bullet ) f_\beta ( \mu - \bullet) ) , \\ 
f_\beta ( x )  := - \beta^{-1}  \log ( e^{ \beta x } +  1) \simeq 
- x_+ , \ \ \beta \to + \infty , 
\end{gathered}
\end{equation}
for smooth $\eta$ localizing to energy intervals contained in $\Delta^{-1}  (-\delta,\delta)$. (Note that $ \vert f_\beta \vert_\alpha $ is uniformly bounded for $ \alpha \leq 1 $ but not for $ \alpha > 1 $.)
The {\em magnetization} is defined by \cite{O52}
\begin{equation}
\label{eq:magn} M_{\beta}(\mu, h): =- \tfrac{3 \sqrt{3}}2 \tfrac{\partial }{\partial h} \Omega_{\beta}(\mu, h).
\end{equation}
and at zero temperature, we can derive from this a \emph{sawtooth approximation}, with $\sigma (y ) :=   y - [y] - \tfrac12,$ as the $ \mathcal O ( h^{\frac12} )$ approximation of \eqref{eq:magn} given as
\begin{equation}
\label{eq:omega3}
\begin{split} 
M_\infty ( \mu, h )  
= \tfrac{1}{\pi} \sigma \left( \tfrac{g ( \mu)} h\right) 
\tfrac{ g ( \mu) }{ g' ( \mu ) } + \mathcal O ( h^{\frac12} ). 
\end{split}  
\end{equation}
This provides a refinement of results found in \cite{SGB94,L11,CM01}.
The remarkable agreement of the different expressions for the magnetization is illustrated in Fig.\ref{Fig:cut-off}: the characteristic sawtooth pattern \eqref{eq:omega3} is compared with the magnetization computed from \eqref{eq:magn}, using either the operator spectrum or the semiclassical limit \eqref{eq:Landau}. 

\smallsection{QHE and self-similarity:} 
One of the striking properties of graphene is the presence of Dirac
points, which has remarkable 
physical and technological implications \cite{WA10}. It turns out
Dirac points are present at $E=0$ for any magnetic flux $h = 2\pi
p/q,$ \cite[Thm. $2$]{B}.
To study transport properties on honeycomb structure (see  \cite{AEG,GS,Pe10,M06}) we consider operator \eqref{eq:Q} with additive disorder
\begin{equation}
\label{eq:rso}
t^h_{\kappa,\omega}= \frac13 \left(\begin{matrix}  -{\kappa V^{(1)}_{\omega}}  && {1+\tau^0+\tau^1}\\ {(1+\tau^0+\tau^1)^*} &&  -{\kappa V^{(2)}_{\omega}} \end{matrix} \right) ,
%(t^h_{\lambda,\omega}u)(\gamma)= \tfrac13 \left[\left(\begin{matrix}  -{\lambda V^{(1)}_{\omega}}  && {1+\tau^0+\tau^1}\\ {(1+\tau^0+\tau^1)^*} &&  -{\lambda V^{(2)}_{\omega}} \end{matrix} \right)u \right](\gamma)
\end{equation} 
where $(V^{1,2}_{\omega(z)})_{z \in \mathbb Z^2}$ are i.i.d. random variables with compactly supported probability distribution and small $\kappa>0$.
For discrete operators $A$ with $\mathbb C^2$-valued kernel $(A(x,y)),$ we define a regularized trace
\begin{equation}
\label{discretetrace}
\widehat{\operatorname{tr}}  A := \lim_{r \rightarrow \infty} \tfrac{1}{\left\lvert B_r(0) \right\rvert} \sum_{\gamma \in \Lambda  \cap B_r(0)} \operatorname{tr}_{\mathbb{C}^{2}} A ( \gamma, \gamma ).
\end{equation}

\begin{equation}
\begin{split}
\label{eq:Landau2}
&\widehat{\operatorname{tr}} f (  t^B) = \tfrac{2q \varepsilon}{3 \sqrt{3} \pi }  \sum_{ n \in \mathbb Z^2 }
f ( z_{n} ( h  ) )+ \mathcal O (\vert f \vert_{ C^\alpha } \varepsilon^\infty)  \\
&\text{ where } z_n(\varepsilon) = v_F\text{sgn}(n) \sqrt{\vert n\varepsilon \vert}+ \mathcal O(\varepsilon) \text{ and }\\
&v_F=3^{3/4}q  \left(3^{q-1}\prod_{j=q+2}^{2q} t_j^{B_0}(\tilde{ k}) \right)^{-1}. 
\end{split}
\end{equation}
Here, $t_j^{B_0}(\tilde{ k})$ is the $j$-th Floquet eigenvalue to $t^{B_0}$ with quasimomentum $\tilde{k}$ where $B_0$ is the magnetic field associated to the flux $h_0=\tfrac{2\pi p}{q}.$ This study is inherently connected with self-similarity in the Hofstadter butterfly, see Fig.\ref{fig:Hofstadter}, and the occurrence of magnetic mini-bands \cite{C14}.
Since $t^B$ is an element of the rotation algebra, so is its Fermi projection $P = \indic_{[z_D,\mu)}(t^B)$ for Fermi energies $\mu$ inside a spectral gap of $t^B$. By \cite{PV1,PV2,R81}, there is $\gamma \in \mathbb Z^2,$ such that
\begin{equation}
\label{eq:Rieffel}
\widehat{\operatorname{tr}}(P) =\frac{2}{3\sqrt{3}} (\gamma_1+\gamma_2 \tfrac{\varepsilon}{2\pi})
\end{equation}
where by \eqref{eq:Landau2} we see that $\gamma=(0,2qn)$ and $n$ is the number of Landau levels between $z_D$ and $\mu.$
Combining \eqref{eq:Landau2} with \eqref{eq:measure} implies the existence of spectral gaps between a finite number of disjoint intervals $\mathcal B_n(h)\ni z_n(h)$ up to some small disorder strength $\kappa_0> 0$. The Hall conductivity, which by universality (see \cite{ASS,BES}) is invariant under weak disorder, is given by St\v{r}eda's formula \cite{S82} as $c_H(\mu):= \tfrac{\partial}{\partial h} \widehat{\operatorname{tr}}(\indic_{[z_D,\mu)}(t^h_{\kappa,\omega})) = \frac{\gamma_2}{2\pi}$ with Fermi energies $\mu$ in the interval $I_n$ between $B_n(h)$ and $B_{n+1}(h)$  \cite[Prop.$1.1$ \& Thm. $4$]{B}. From \eqref{eq:Landau2} and \eqref{eq:Rieffel} we then find 
\begin{equation}
\begin{split}
\label{eq:Hall}
c_{H}(\mu) &= \begin{cases} & \frac{(2n+1)q}{2\pi}, \ \mu \in I_n,\ n \ge 0 \\
&\frac{(2n-1)q}{2\pi}, \ \mu \in I_{n-1}, \ n \le 0.
\end{cases}
\end{split}
\end{equation} 
This expression is only valid for Fermi energies close to the conical point. The Hall conductivity for arbitrary Fermi energies is far more intricate, see Fig.\ref{fig:Hofstadter}, \cite{Z,AEG}.

\smallsection{Metal/insulator transition:} 
The Hall conductivity allows us also to analyze transport properties of $t^h_{\kappa,\omega}.$ Transport in disordered media at energy $E$ is measured by transport coefficients $\beta^h_{\kappa}(E)$ \cite{GK,GK2,GK3,GKS,GK6}. This quantity allows us to define two complementary energy regions, the \emph{insulator region} $\Sigma_{\kappa}^{h,\text{DL}} = \left\{ E \in \mathbb R; \beta^h_{ \kappa}(E)=0 \right\}$
and the \emph{metallic transport region} $\Sigma_{\kappa}^{h,\text{DD}} = \left\{ E \in \mathbb R; \beta^h_{ \kappa}(E)>0 \right\}.$
Energies $E \in \Sigma_{\kappa}^{h,\text{DD}} $ at which the transport coefficient $\beta^h_{\kappa}$ jumps from zero to a non-zero value are called \emph{mobility edges},
while energies $E \in  \Sigma_{\kappa}^{h,\text{DL}}(H^h_{\kappa,\omega})$, that also belong to the spectrum of \eqref{eq:rso}, are eigenvalues of finite multiplicity with exponentially decaying eigenfunctions \emph{(Anderson localization)}.  
From the jumps of the Hall conductivity, we conclude \cite[Thm.
$1$]{B} that there exist mobility edges $E$ close to each Landau level
with non-trivial transport $\beta^h_{\lambda}(E)\ge 1/4.$ In contrast
to this, we show by verifying the starting criteria of the multi-scale
analysis \cite{B,FS83,GK2} that the spectral gaps between the Landau
levels can only be filled with spectrum belonging to the insulating region \cite[Prop.\@$5.5$]{B} in which the operator \eqref{eq:rso} therefore exhibits Anderson localization.

\smallsection{Acknowledgements} S.B.~gratefully acknowledges support by
the UK Engineering and Physical Sciences Research Council (EPSRC)
grant EP/L016516/1 for the University of Cambridge Centre for Doctoral
Training, the Cambridge Centre for Analysis. R.H., S.J., and M.Z.~were partially supported
by the National Science Foundation under the grants DMS-1800689,
 1500852 and 1901462. %The authors would also like to thank Hari
%Manoharan for introducing us to molecular graphene and for allowing us to use Figure \ref{Fig:DOS}(b).

\end{document}